\documentclass[aip,jcp,reprint]{revtex4-1}

\usepackage{dcolumn}
\usepackage{bm}
\usepackage[T1]{fontenc}
\usepackage[utf8]{inputenc}
\usepackage{amsfonts}
\usepackage{amsmath}
\usepackage{amssymb}
\usepackage{siunitx}
\usepackage[justification=raggedright]{caption}
\usepackage{subcaption}
\usepackage{grffile}
\usepackage{graphicx}
\usepackage{hyperref}
\usepackage{color}
\usepackage{ulem}

\begin{document}


\title{Non-conformal coarse-grained potentials for water} 



\author{Tonalli~Rodr\'{\i}guez-L\'{o}pez}
\email[E-mail: ]{xclassicx@gmail.com}
\affiliation{
Departamento de F\'{i}sica, 
Universidad Aut\'onoma Metropolitana, Iztapalapa, Apdo 55 534, M\'exico DF 09340, M\'exico.
}
\affiliation{
Department of Chemistry \& Waterloo Institute of Nanotechnology, 
University of Waterloo, 
200 University Avenue West, Waterloo, Ontario, Canada N2L\,3G1
}

\author{Yuriy Khalak}
\email[E-mail: ]{khalak.yura@gmail.com}
\affiliation{
Department of Mathematics and Computer Science \& Institute for Complex Molecular Systems, 
Eindhoven University of Technology, P.O. Box 513, MetaForum, 
5600 MB Eindhoven, the Netherlands
}

\author{Mikko Karttunen}
\email[E-mail:  ]{mkarttu@gmail.com}
\affiliation{
Department of Mathematics and Computer Science \& Institute for Complex Molecular Systems, 
Eindhoven University of Technology, P.O. Box 513, MetaForum, 
5600 MB Eindhoven, the Netherlands
}
\affiliation{
Department of Chemistry, Western University,
1151 Richmond Street, London, Ontario, Canada N6A\,5B7
}
\affiliation{
Department of Applied Mathematics, Western University,
1151 Richmond Street, London, Ontario, Canada N6A\,5B7
}

\date{\today}

\begin{abstract}

Water is a notoriously difficult substance to model both accurately and efficiently. 
Here, we focus on descriptions with a single coarse-grained particle per molecule using the so-called Approximate Non-Conformal (ANC) and generalized Stockmayer potentials as the starting points. 
They are fitted using the {radial distribution function} and the {liquid-gas density profile} of the atomistic SPC/E model by downhill simplex optimization. 
We compare the results with monatomic water (mW), ELBA, as well as with direct Iterative Boltzmann Inversion (IBI) of SPC/E. 
The results show that symmetrical potentials result in non-transferable models, that is, they need to be reparametrized for new state-points. 
This indicates that transferability may require more complex models.
Furthermore, the results also show that the addition of a point dipole is not sufficient to make the potentials accurate and transferable to different temperatures (300\,K-500\,K) and pressures without an appropriate choice of properties as targets during model optimization.

\end{abstract}


\maketitle 

\section{Introduction} \label{section:introduction}

Water is arguably the most important and studied substance in the world. It covers approximately 70\% of the Earth's surface and is vital for all known forms of life. In particular, at the microbiological scale, membranes that provide the container and selective protection for all cells, as well as proteins, peptides, and drugs, all require an aqueous solution to function.\cite{Chaplin2006,Ball2008}

Despite being the most studied substance, our understanding of water is far from comprehensive. Its physical behavior\cite{Kuehne2013,Skinner_2014,Palmer2014,Titantah_2015} and  interactions with biological molecules\cite{Thirumalai2012,Bellissent-Funel2016} keep providing surprises. Developing computational and theoretical models for water has proven to be extremely demanding and remains a topic of intense research\cite{Guillot2002a,Wu2010a,Mao2012,Izadi_2014,Tainter2015,Pethes_2017,Steinczinger_2017}.

The focus of this work is on coarse-grained (CG) single-particle models of water. The general goal of coarse-graining is to reduce the number of degrees of freedom allowing for significantly longer simulation times and larger systems yet retaining the essential physical properties\cite{Murtola2009d,Riniker2012d,Ingolfsson2013}. There is no single approach to coarse-graining and a large number of CG models with unique strengths and weaknesses exist\cite{Murtola2009d,Riniker2012d,Ingolfsson2013,Lazaridis_2014}. Apart from implicit solvent models, each of the approaches has their own model(s) for water\cite{Hadley2012,Darre2012,Lazaridis_2014}. Here, we operate at a modest level of coarse-graining to be able to make direct comparisons with the underlying atomistic description.

The reduction in degrees of freedom and averaging over properties such as electrostatic interactions and hydrogen bonding can sometimes lead to unwanted and surprising effects.
{
Observed effects include freezing of water at higher temperatures\cite{Marrink2007} (this occurs with the original water of the very popular MARTINI model; water freezes at $290 \pm 5$\,K but the polarizable MARTINI water model later corrected this to lower temperatures between 280 and 285\,K\cite{Yesylevskyy2010a}),
freezing and ordering of CG water close to surfaces\cite{Bennun2007a,Xing2008,Habibi2014} or the tradeoff between reproducing the structure factor and tetrahedral packing with single water models\cite{Hadley2012,Wang2009}.
Each of the coarse-grained (as well as atomistic) water models have their strengths and weaknesses, and it is important to pay attention that the chosen model is approriate for the intended range of physical parameters such as temperature or pressure.  Darre \textit{et al.}, and Hadley and McCabe provide recent reviews of coarse-grained water models\cite{Darre2012,Hadley2012}.
}

To overcome some of these issues,  we introduce three customized models for CG water based on the Approximate Non-Conformal (ANC) potential\cite{F.del.Rio.&.J.E.Ramos.&.I.A.McLure-1998}. These new models, called ANC-1G, ANC-2G and GSM, will be detailed below. We characterize them and compare their behavior to some of the well-known CG water models, namely monatomic Water (mW)\cite{Molinero2009} and the electrostatic-based (ELBA) model\cite{Orsi2011,Orsi2013}. As an all-atom reference model for water, we use the extended simple point charge (SPC/E) model\cite{Berendsen1987} and apply the iterative Boltzmann Inversion (IBI) method\cite{Reith2003} to it to obtain a CG reference. 

Majority of modern water (and biomolecular) models, whether coarse-grained or atomistic, rely on the Lennard-Jones potential. The motivation for diverting from the usual Lennard-Jones potential-based approach in favor of a non-conformal model is that liquid water has already long ago been shown to be poorly described by van der Waals' law of corresponding states\cite{van1913law} resulting in models that only qualitatively reproduce density and compressibility dependence on pressure\cite{Watson1943, Dong1986}. As shown by Pitzer already in 1939, the law of corresponding states holds exactly only for fluids with conformal potentials, that is, potentials that remain invariant under energy and distance scaling like Lennard-Jones. To describe interactions more realistically beyond Lennard-Jones, extensions and generalizations of non-conformal models involving more parameters especially in the context of colloids have been developed, see e.g. Refs.~\onlinecite{Noro_2000,Foffi_2007}.

\section{Models and Methods} \label{section:models and method}
\subsection{Models}
\subsubsection{Atomistic reference model}

The reference model, SPC/E water\cite{Berendsen1987},  is characterized by three point masses with oxygen-hydrogen distance of \SI{1.0}{\angstrom} and the HOH angle equal to the tetrahedral angle \ang{109.47}, with charges on the oxygen and hydrogen equal to \SI{-0.8476}{\elementarycharge} and \SI{+0.4238}{\elementarycharge}, respectively, and with the Lennard-Jones parameters of oxygen-oxygen interaction set to $\epsilon_\mathrm{LJ} = 0.1535$ kcal/mol and $\sigma_\mathrm{LJ} = \SI{3.166}{\angstrom}$. We chose SPC/E  since it is commonly used and reproduces properties such as isothermal compressibility and critical point temperature better than most other popular three-site models\citep{Vega2011}. While a four-site model like TIP4P/2005\citep{Abascal2005} does yield more accurate liquid densities\citep{Abascal2005,Vega2011}, it is less frequently used in large-scale simulations.

\subsubsection{New CG parameterizations}

Using the SPC/E model as a reference, we parametrized three new coarse-grained models, ANC-1G, ANC-2G and GSM. In them, water molecules are represented as single particles that interact through either isotropic (ANC-1G and ANC-2G) or anisotropic (GSM) interactions. All three are based on the original ANC pair potential of del Rio et al.\cite{F.del.Rio.&.J.E.Ramos.&.I.A.McLure-1998}

{
In the procedure below, we fit the potential twice using the the radial distribution function as the first target and the liquid-gas density profile as the second. This allows us to determine which target property provides a better model. Structure based fitting as the sole target is used in methods such as  the iterative Boltzmann Inversion (IBI)\cite{Reith2003}. The hope is that adding a second target will help to obtain a better model; density was chosen because it is a commonly used metric in the development of atomistic water models.  We would like to note that this approach is different from using multiple simultaneous targets as is used with, for example, the mW\cite{Molinero2009} and ELBA\cite{Orsi2011,Orsi2013} models. Fitting to multiple properties at once produces models that, at least in principle, approximate more properties at once, but do not necessarily exactly match any. Thus, fitting separate models to individual properties and fitting a single model to multiple properties simultaneously are two different philosophies. The results here provide a comparison between the two approaches. 
}

The ANC potential retains the length  and energy  parameters from Lennard-Jones, and enforces non-conformality through a third parameter called softness\cite{F.del.Rio.&.J.E.Ramos.&.I.A.McLure-1998}.
The ANC pair potential is given by\cite{F.del.Rio.&.J.E.Ramos.&.I.A.McLure-1998}
\begin{equation}
\phi_\mathrm{anc} \equiv  
\phi_{\rm{anc}} \left( r_{ij}; \delta_{\rm{m}}, \epsilon, s \right)
\! = \!
\epsilon
\left[ 
   \left(\frac{\delta_{\rm{m}}}{\zeta_{ij}}\right)^{12}\!-\!2
   \left(\frac{\delta_{\rm{m}}}{\zeta_{ij}}\right)^{6}
\right], 
\label{eq:upair_anc}
\end{equation}
\begin{equation}
\rm{where\:}
 \zeta_{ij}^{3} = \delta_{\rm{m}}^3 + \left[ {r_{ij}^3  - \delta_{\rm{m}}^3} \right] / s ,
 \label{eq:upair_anc_zeta}
\end{equation}
and $r_{ij}$ denotes the relative distance between the \textit{i}th and \textit{j}th particle, $(\delta_{\rm{m}})$ length scale, $(\epsilon)$ the depth of the potential well, and non-conformality is enforced by the softness of the potential $(s)$. Softness measures the ratio between of the slope and that of the reference potential, in this case the Lennard-Jones. The reference potential is recovered at $s = 1$.
It has been shown to provide an excellent approximation for many gases\cite{I.A.McLure.&.J.E.Ramos.&.F.del.Rio-1999,F.del.Rio.&.J.E.Ramos.&.I.A.McLure-1999,J.E.Ramos.&.F.del.Rio.&.I.A.McLure-2000,J.E.Ramos.&.F.del.Rio.&.I.A.McLure-2001}. Furthermore, it has an analytical expression for the second virial coefficient.\cite{Gonzalez-Calderon2015}

As a new step, we  add Gaussian functions to the ANC potential. This is motivated by the fact that structure-based inversion methods, such as Inverse Monte Carlo, Inverse Boltzmann and Force Matching, show that coarse-graining from SPC/E and TIP4P atomistic water models yields effective potentials with two minima.\cite{Lyubartsev2003,Wang2009,Ru2009} A straightforward way to implement this is to add a Gaussian function to the ANC potential.  The Gaussian contribution can be given by
\begin{equation}
\phi_{\rm{g}} \left( r_{ij}; h,  p, q \right) =
h \exp
\left[ 
  -\frac{\left(r_{ij} - p \right)^2}{2 q^2} 
\right]
\label{eq:gaussian}
\end{equation}
with peak height $h$, peak position $p$, and a standard deviation $q$. By adding one and two Gaussian functions to the ANC potential (Eq.~\ref{eq:upair_anc}), we obtained the ANC-1G and ANC-2G potentials, respectively. 
While some previous studies have employed as many as four Gaussian functions\cite{Barraz2009}, we determined that using two is sufficient in order not to introduce excess free parameters.

The ANC, ANC-1G and ANC-2G are still symmetric. Another way to extend the ANC potential to capture more molecular properties is to introduce orientational dependence. In atomistic water models, this is usually achieved by placing opposite atomic partial charges on hydrogens and oxygens to induce intermolecular hydrogen bonding, and at the coarse-grained level the Polarizable Martini\cite{Yesylevskyy2010a}, mW\cite{Molinero2009} and the Mercedes-Benz\cite{Dias2009} models have orientational dependence. In the latter two, the approach is inspired by the models for silicon which also exhibit tetrahedral order. Both the mW and Mercedes-Benz model also display the density anomaly of water. The polarizable Martini model, albeit at a higher level of coarse-graining, uses a fluctuating dipole thus resembling the GSM approach here. We use the GSM\cite{E.Avalos.&.F.del.Rio.&.S.Lago-2005} potential, which combines a point dipole from the Stockmayer\cite{Stockmayer1941} potential with the ANC potential (Eq.~\ref{eq:upair_anc}) without any Gaussian functions:
\begin{equation}
\phi_{\rm{gsm}}\left(r_{ij}, \Omega_{ij};  \delta_{\rm{m}}, \epsilon, \mu \right)
=
\phi_{\rm{anc}}
+
\phi_{\rm{dd}}\left(r_{ij},\Omega_{ij};\mu \right),
\label{eq:upair_gsm}
\end{equation}
where $\phi_{\rm{dd}}$ is the dipole-dipole interaction given as
$$
\phi_{\rm{dd}}\left(r_{ij},\Omega_{ij};\mu \right)
=
\frac{1}{4 \pi \varepsilon_{0}}
\left[
  \frac{ \vec{\mu}_{i} \cdot \vec{\mu}_{j} }{ r_{ij}^{3} }
  -3
  \frac{ \left( \vec{r}_{ij} \cdot \vec{\mu}_{i} \right)\left( \vec{r}_{ij} \cdot \vec{\mu}_{j} \right) }{ r_{ij}^{5} }
\right],
$$
with $\mu$ being the dipole strength, $\varepsilon_{0}$ the permittivity of vacuum, and $\Omega_{ij}$ stands for the set of angles that defines the relative orientation between molecules $i$ and $j$. In this case, the Stockmayer potential and dipolar hard spheres are obtained from Eq.\,\eqref{eq:upair_gsm} with $s=1$ and $s \to 0$, respectively.

\subsubsection{Other CG models}

Since we want to assess transferrability to different state points,
we also studied some of the most popular coarse-grained water models to compare with the new ANC and GSM potentials, namely, monatomic water (mW)\cite{Molinero2009} water and the electrostatic-based (ELBA) model\cite{Orsi2011,Orsi2013}.

The mW model\cite{Molinero2009} is an adaptation of the Stillinger-Weber silicon potential\cite{Stillinger1985} that favors a tetrahedral coordination of molecules. The mW model was parameterized to reproduce the experimental melting temperature of hexagonal ice as well as the density and enthalpy of vaporization of liquid water at ambient conditions\cite{Molinero2009}. In the mW model, each water molecule is mapped to one bead that interacts through both a two-body and a three-body potential, described by
 \begin{equation} 
   E = \sum\limits_i {\sum\limits_{j > i} {{\phi _2}\left( {{r_{ij}}} \right)} }  + \sum {\sum {\sum {{\phi _3}\left( {{r_{ij}},{r_{ik}},{\theta _{ijk}}} \right)} } } ,
\end{equation}
where
$$
{\phi _2}\left( {{r_{ij}}} \right) = A\varepsilon \left[ {B{{\left( {\frac{\sigma }{{{r_{ij}}}}} \right)}^4} - 1} \right]
\exp \left( {\frac{{\gamma \sigma }}{{{r_{ij}} - a\sigma }}} \right)
$$
and
\begin{gather}
  \begin{aligned}
  {\phi _3}\left( {{r_{ij}},{r_{ik}},{\theta _{ijk}}} \right) & = \lambda \varepsilon {\left[ {\cos {\theta _{ijk}} - \cos {\theta _0}} \right]^2}
  \nonumber \\
  & \times \exp \left( {\frac{{\gamma \sigma }}{{{r_{ij}} - a\sigma }}} \right)\exp \left( {\frac{{\gamma \sigma }}{{{r_{ik}} - a\sigma }}} \right), \nonumber
\end{aligned}
\end{gather}
where $A=7.049556277$ and $B=0.6022245584$. The rest of the parameters are $\theta = \ang{109.47}$, $\epsilon = 6.189$ kcal/mol, $\sigma = \SI{2.3925}{\angstrom}$, $a = 1.8$, $ \lambda = 23.15 $, and $ \gamma = 1.2 $.\cite{Molinero2009}

Lastly, we also included the ELBA model\cite{Orsi2011,Orsi2014} originally developed as a solvent for lipid membranes with the aim of reducing the computational cost. The ELBA model describes a water molecule as a single CG bead embedded with a point dipole. The potential energy for a pair of CG beads is the sum of Lennard-Jones and dipole interactions, both terms in a shifted-force form and are given by
\begin{gather}
\begin{aligned}
  \phi _{\rm{sf - lj}}  =  & 4\epsilon
  \left\{ 
    \left[
           \left( \frac{\sigma }{r_{ij}} \right)^{12} 
       -   \left( \frac{\sigma }{r_{ij}} \right)^6
    \right]
  \right.
 \\
  & +  \left[
        6\left( \frac{\sigma }{r_c} \right)^{12} 
        - 3\left( \frac{\sigma }{r_c} \right)^6
      \right]
      \times \left( \frac{r_{ij}}{r_c} \right)^2
  \\ 
 & - \left. 7\left( \frac{\sigma }{r_c} \right)^{12}
  + 4\left( \frac{\sigma }{r_c} \right)^6 
\right\} 
\end{aligned}
\end{gather}
and
\begin{gather}
\begin{aligned}
{\phi _{\rm{sf - dd}}} & = 
 \frac{1}{{4\pi {\varepsilon _0}}}
 \left[ 
  {1 - 4{{\left( {\frac{{{r_{ij}}}}{{{r_c}}}} \right)}^3} + 3{{\left( {\frac{{{r_{ij}}}}{{{r_c}}}} \right)}^4}} 
 \right] \nonumber \\
& \times 
 \left[ 
   {\frac{{\left( {{{\vec \mu }_i} \cdot {{\vec \mu }_j}} \right)}}{{r_{ij}^3}} -
     \frac{{3\left( {{{\vec r}_{ij}} \cdot {{\vec \mu }_i}} \right)\left( {{{\vec r}_{ij}} \cdot {{\vec \mu }_j}} \right)}}{{r_{ij}^5}}} 
\right]
\end{aligned}
\end{gather}
where the set of parameters are $\epsilon = 0.55 ~\rm{kcal}~\rm{mol}^{-1}$, $\sigma = \SI{3.05}{\angstrom}$, $\mu = 2.6 ~\rm{D}$, and $r_{\rm{c}} = \SI{12}{\angstrom}$. This model was originally parametrized to reproduce bulk density and diffusion coefficient of liquid water at \SI{30}{\degreeCelsius} and {1}{ atm}\cite{Orsi2011}.

\subsection{Model Optimization}

The model parameters for both the ANC and GSM models were optimized to best reproduce the behavior of the SPC/E reference system. Parametrizations were made using both bulk and slab systems, using the radial distribution function and the density profile, respectively, as the target properties. The ANC parametrizations, however, became unstable at the intermediate optimization parameters for the slab geometry.
The general procedure for potential optimization has been previously described elsewhere\cite{Ru2009} and implemented in the Versatile Object-Oriented Toolkit for Coarse-graining Applications (VOTCA) package\cite{Ru2009,Mashayak2015} which was used here. Here, the focus is on the selection of the potential parameters during the potential update step.

Due to limited sampling, results of MD simulations contain noise, which can lead to significant errors in numerically determined gradients of the target properties. Therefore, gradient based function optimization methods are a poor choice for optimizing model parameters. To circumvent this limitation, we used the downhill simplex algorithm\cite{Nelder1964} to optimize all model parameters. It is a deterministic optimization procedure designed to minimize an objective (or penalty) function of $n + 1$ variables, $y(\mathbf{x}_1, \mathbf{x}_2, \ldots, \mathbf{x}_{n+1})$. This method requires only function evaluations, not derivatives. The idea is to employ a moving simplex in the \textit{n}-dimensional parameter space to surround the optimal point and then shrink the simplex until its dimensions reach a specified error tolerance. A simplex
is a geometrical figure consisting, in $n$ dimensions, of $n + 1$ vertices $\mathbf{x}_1, \mathbf{x}_2, \ldots, \mathbf{x}_{n+1}$ connected by straight lines and bounded by polygonal faces. It is transformed successively using basic operations such as reflection, expansion, contraction, or reduction in order to move towards the global minimum, i.e., to minimize the objective function\cite{Nelder1964}.
{
In the context coarse-graining, the use of simplex optimization was pioneered in the M{\"u}ller-Plathe group for coarse-graining of polymers~\cite{Faller1999,Meyer2000,Reith2001}. It is currently also used in speeding up optimization of paremeters for complex biomolecular force fields such as Amber and others\cite{Betz2014,Mayne2013}. 
}

To obtain a reference for the CG pair potential and to verify consistency, we also employed the IBI method\cite{Reith2003} with the atomistic structure of the bulk system as the target property.
Both the simplex procedure and IBI were performed with the VOTCA package.\cite{Ru2009,Mashayak2015}

\subsection{Simulation details}

\begin{figure}[ht]
\includegraphics[width=\columnwidth]{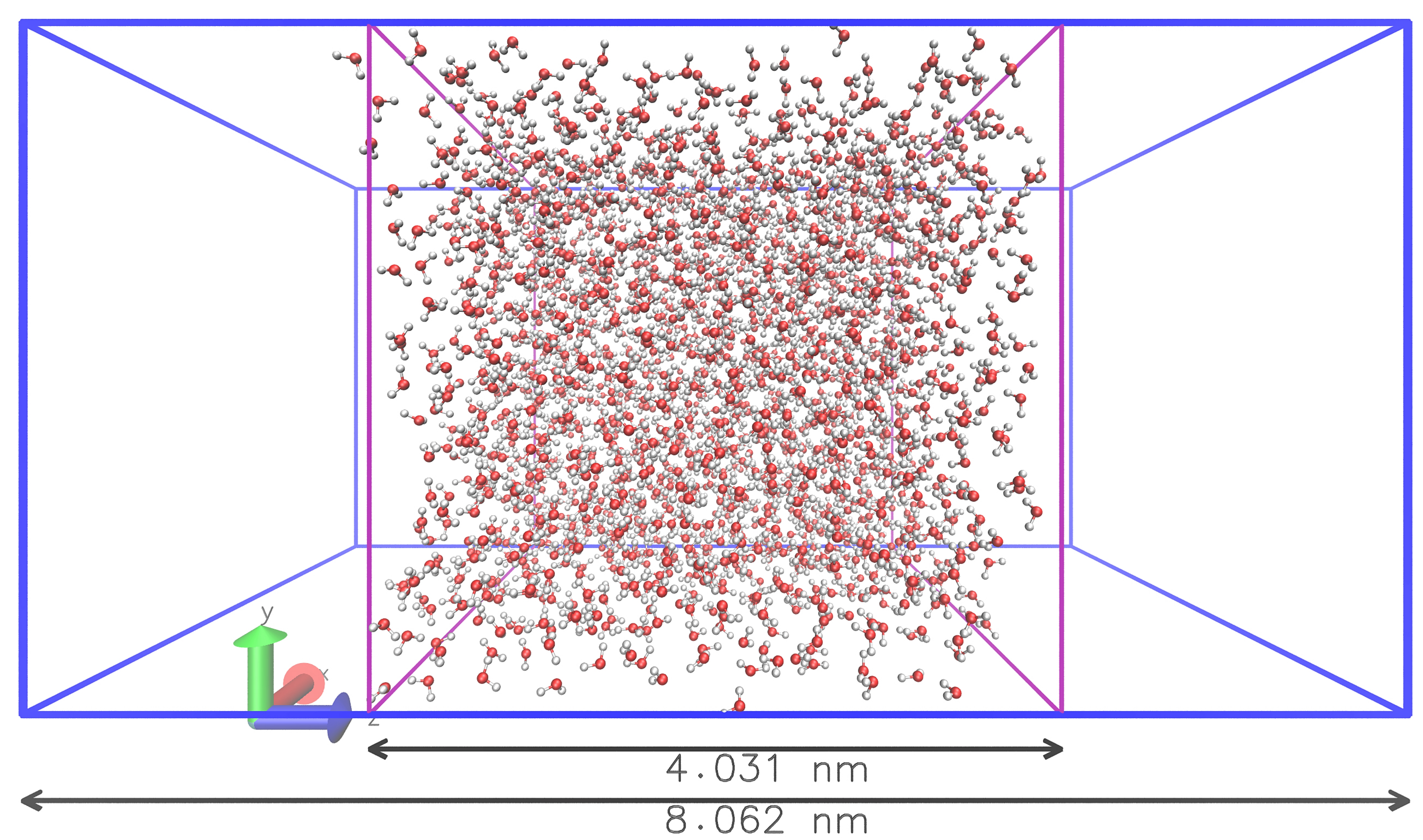}
\caption{Configuraton of the slab system. Bulk water is placed into the center and spreads out along the \textit{Z}-axis, forming regions of smaller density, sometimes evaporating into the surrounding vacuum. Figure produced with VMD\cite{Humphrey1996}.}
\label{fig:slab_conf}
\end{figure}

For the reference system, each SPC/E water molecule was mapped onto its center of mass to represent a one-bead coarse-grained water molecule. Each SPC/E system consisted of $2,180$ water molecules.
A cubic box of side $l_{b} = \SI{40.31}{\angstrom}$ was used for the bulk system, and a rectangular box obtained by expanding the cubic box by a factor of two in the \textit{z}-direction $(l_{s} = 2 l_{b})$ for the slab system, generating a slab of liquid water surrounded by vapor on both sides (Fig.~\ref{fig:slab_conf}). Equations of motions were integrated with the velocity-Verlet algorithm\cite{Swope1982} and a time step of \SI{2}{\femto\second}.
Both reference and CG simulations were simulated in the canonical ensemble (constant number of molecules $N$, constant volume $V$ and constant temperature $T$) using the Nos\'{e}-Hoover thermostat\cite{Nose1984,Hoover1985} with $T = \SI{300}{\kelvin}$ and a relaxation time of \SI{200}{\femto\second}, except for CG liquid-gas equilibrium simulations, where a relaxation time of \SI{500}{\femto\second} was used. In the case of the GSM model, the NVT/sphere
integrator was used for dipoles in order to update the orientation of the dipole moment and to apply the  Nos\'{e}-Hoover thermostat to its rotational degrees of freedom. This was necessary to prevent large amounts of energy from accumulating in rotational degrees of freedom while translational degrees of freedom experience a much smaller effective temperature. All MD simulations were performed with the LAMMPS package\cite{Plimpton1995} modified to permit simulations using the ANC potential.

Atomistic reference simulations were run for \SI{10}{\nano\second}, bond lengths and angles were constrained using the SHAKE algorithm\cite{Ryckaert1977},  long-range electrostatic interactions were handled with the particle-particle particle-mesh method\cite{hockney1988computer,Pollock1996}, and the van der Waals interactions were cut off at \SI{10}{\angstrom}. For the case of point dipoles, long-range interactions were handled with the ewald/disp solver.\cite{Toukmaji2000}

About $200-400$ simplex iterations were performed for each of CG pair potentials studied. The number of iterations depends on the selection of the initial set of parameters and can not be determined \textit{a priori}. For faster (wall time) convergence of the parameters, sampling times were kept short, similar to other works using the VOTCA package for coarse-graining\cite{Ru2009,Mashayak2015}.
During every iteration, starting from the same initial configuration, a damped dynamics energy minimization method, called quick-min\cite{Sheppard2008}, was performed, followed by CG simulations for \SI{100}{\pico\second} with the first \SI{50}{\pico\second} ignored during analysis. Radial distribution functions, density profiles, and pair potentials (Figs.~\ref{fig:bulk} and \ref{fig:slab}) were obtained from these simulations.

\subsection{Measurement of Densities in Liquid-Gas Equilibria}
To measure liquid and gas densities as a function of temperature, we performed \SI{1}{\nano\second} simulations using the slab geometry (Fig.~\ref{fig:slab_conf}). The systems were divided into equally sized sections along the \textit{z}-axis for analysis and the average density of each section $\rho(z)$ was computed. A periodic function
\begin{equation}
\rho_\mathrm{fit}(z)= \rho_g + \sum_{n=-2}^{2}{f(z\bmod (l_b), n l_b)}
\label{eq:periodic}
\end{equation}
composed of a sum of logistic functions
\begin{equation}
f(z, n l_b)= \frac{\rho_l-\rho_g}{1+e^{-k(z-z_0+n l_b)}} - \frac{\rho_l-\rho_g}{1+e^{-k(z-z_1+n l_b)}}
\label{eq:logistic}
\end{equation}
was fitted to reproduce the densities from the simulations using the non-linear least squares curve fitting algorithm of SciPy\cite{scipy}. Each pair of logistic functions (Eq.~\ref{eq:logistic}) represents a region of liquid water of density $\rho_l$ surrounded by water vapor of density $\rho_g$ located in a periodic cell $n$ with cell width $l_b$ along the z-axis. The transition regions between the two states are characterized by a decay parameter $k$ and  are centered on $z_0$ and $z_1$. Use of a periodic function with contributions from multiple periodic images of the same liquid region (Eq.~\ref{eq:periodic}) is required to account for the possibility of a liquid-gas transition occurring near the periodic boundaries and to ensure a smooth curvature of the fitted function in the region.

\section{Results and discussion} \label{section:results}

In this work we employed two kinds of interactions: 1) isotropic, Eq.\,\eqref{eq:upair_anc},  2) anisotropic via inclusion of a point-dipole interaction, Eq.~\eqref{eq:upair_gsm}. In both cases, parameters were obtained from the radial distribution functions and density profiles. We present first the results obtained from bulk systems, and then those obtained from slab geometry. To test transferability, the parameters were also employed in simulations over a temperature range  from \SIrange{300}{550}{\kelvin}.

\subsection{Bulk water system}
\begin{figure}[ht]
  \includegraphics[width=\columnwidth]{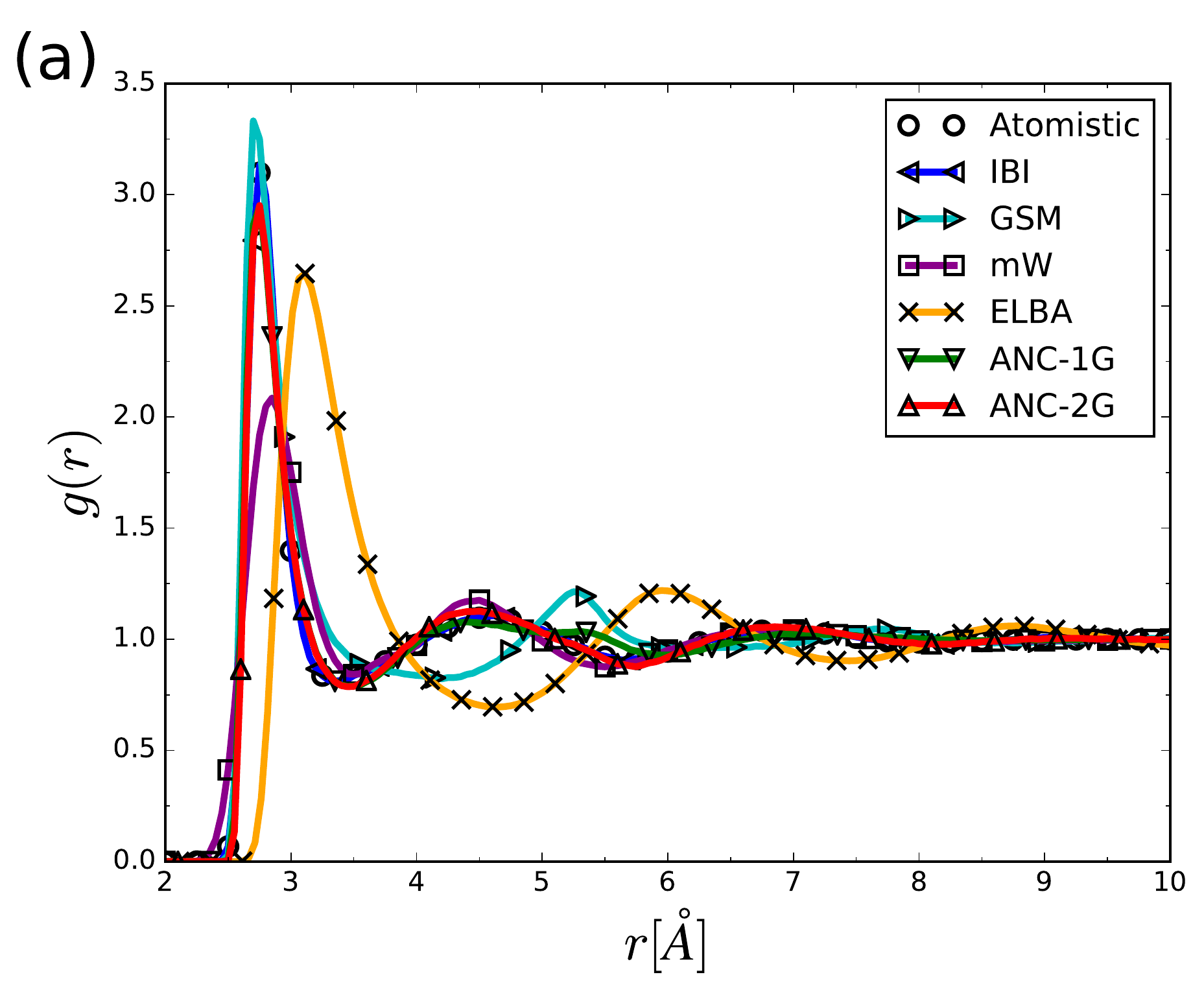} \\
  \includegraphics[width=\columnwidth]{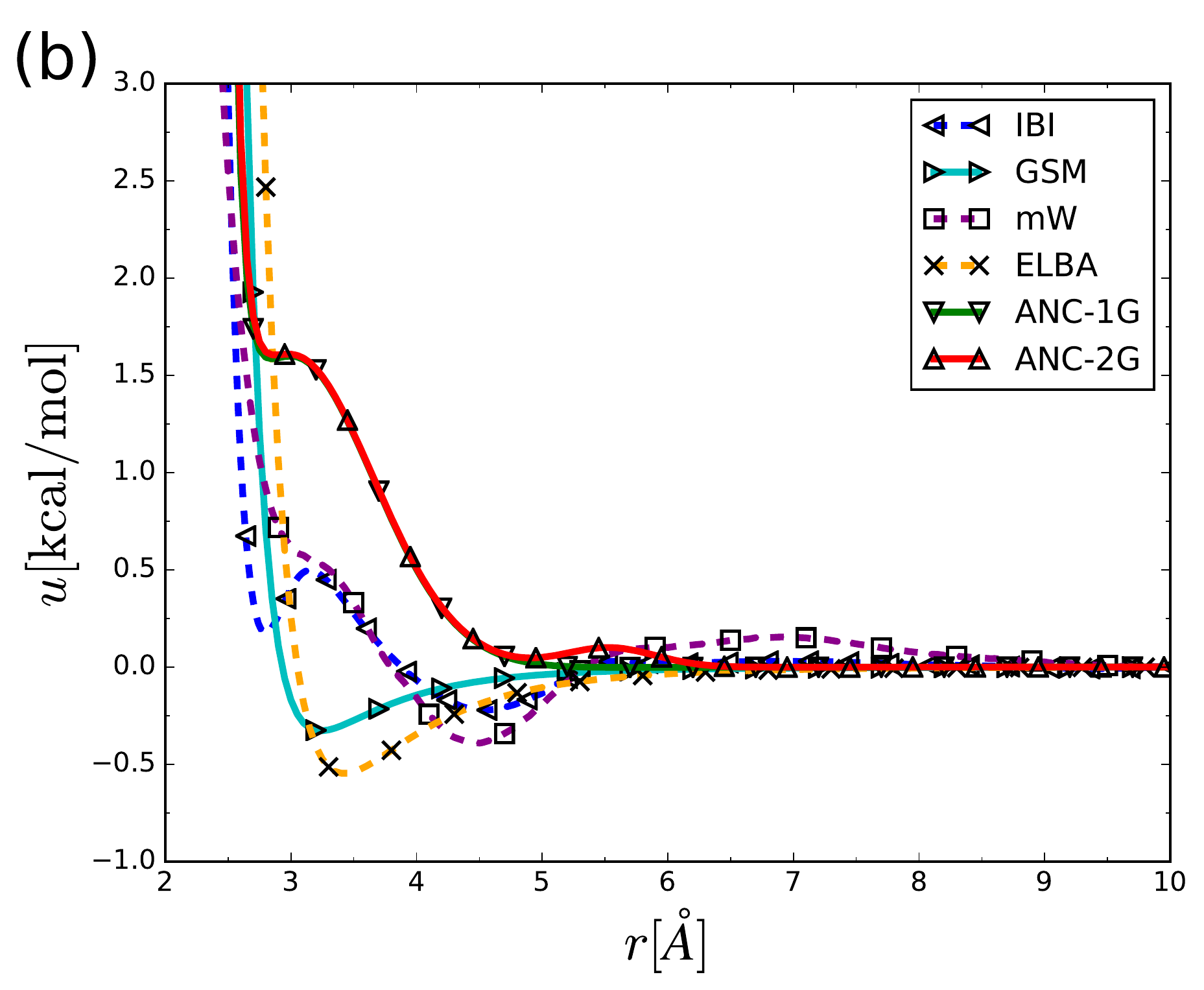}
\caption{Radial distribution (a) and pair potential (b) of ANC-1G, ANC-2G, and GSM potentials following Simplex optimization of the bulk water system using the radial distribution function as the target property.
  Results from iterative Boltzmann inversion (IBI) method and the atomistic reference\cite{Berendsen1987} are also shown. These are further compared to previously published models\cite{Molinero2009,Orsi2011,Orsi2013}. The pair potential shown for GSM does not include the dipole contributions, while the pair potential for mW is obtained from an IBI fit to the mW model and shows an equivalent potential lacking 3-body interactions.
  In general all dashed pair potentials are results of IBI.}
\label{fig:bulk}
\end{figure}

The radial distribution functions $g(r)$ and the pair potentials $u(r)$ obtained from the simplex procedure are shown in Figs.~\ref{fig:bulk}a and b, respectively, for both isotropic and anisotropic interactions. The IBI results, the atomistic reference, as well as the results for
mW and ELBA models are shown for comparison.

Figure~\ref{fig:bulk}a shows that using the ANC potential with a single Gaussian function (ANC-1G) yields an additional unphysical peak in $g(r)$ near \SI{5.3}{\angstrom}. We were able to correct this by applying a second Gaussian function to the potential (ANC-2G). The resulting $g(r)$ follows very closely the $g(r)$ from IBI and the underlying atomistic reference (SPC/E) model. This is similar to previous findings\cite{Lyubartsev2003,Wang2009,Ru2009} showing that a spherically symmetrical single particle potential needs to have two wells to recover atomistic water structure. As Fig.~\ref{fig:bulk}b shows, the IBI procedure produces two potential wells, where their maxima correspond to radii of the first and second hydration shells. Both of the ANC models are purely repulsive, as the added Gaussian functions fill up the minimum of the base ANC potential, Eq.\,\eqref{eq:upair_anc}. Meanwhile the IBI effective potential does have an attractive region. Therefore, it is possible that better parameters for ANC-1G and ANC-2G exist but are unreachable by the simplex optimization procedure or require thermodynamic properties as targets in addition to structural ones.

The $g(r)$ of the GSM potential, on the other hand, exhibited a significantly shifted narrow second peak at \SI{5.3}{\angstrom} and a very broad and flat first minimum at \SIrange{3.3}{4.5}{\angstrom}. The first peak, however, closely follows the reference system.
These discrepancies indicate that adding a point dipole to a single particle water model is not sufficient to make it reproduce atomistic structure. Furthermore, visual inspection of the trajectory revealed that particles described by the GSM potential were not percolating very far.

As the above shows, addition of a simple dipole moment is insufficient to capture the effects of tetrahedral hydrogen bonding of water. Addition of a quadrupole moment may be required, but that was not done in the current study. The mW water model mimicks this effect with a three-body term instead.
With the exception of a slightly wider first peak, the mW water reproduces the reference $g(r)$ very well. Although it does not have a second minimum in its effective potential (but has a three-body term), the repulsive force drops significantly near \SI{3.2}{\angstrom}, the region of the first minimum in the $g(r)$ of the IBI and the reference atomistic potentials. This is similar to the behavior of ANC-1G and ANC-2G in the same region, Fig.~\ref{fig:bulk}b.

\subsection{Slab system}

As an alternative means of model parametrization, we tried to fit to the density of a slab configuration at \SI{300}{\kelvin}. As discussed by Ismail et al.\cite{Ismail_2006}, density profile can be used as an order parameter in a system containing liquid-vapor interfaces. Unlike Ismail et al. who compared surface tensions in different atomistic water models and studied capillary waves, we restrict ourselves to the density profile only and use it to obtain the vapor-liquid phase diagram (the next section). However, this approach failed for the ANC-1G and ANC-2G potentials, as during optimization the simulation code could not handle the intermediate parameters.
This is appears to be due to the repulsive nature of those potentials. In addition, as shown by  Ismail et al.\cite{Ismail_2006}, even the atomistic models underestimate surface tension. In that light, the failure of the  ANC-1G and ANC-2G potentials is not a surprise. Improvements are a topic of a future study.

\begin{figure}[ht]
\begin{flushleft}
    \end{flushleft}
\vspace*{-0.3cm}
    \includegraphics[width=\columnwidth]{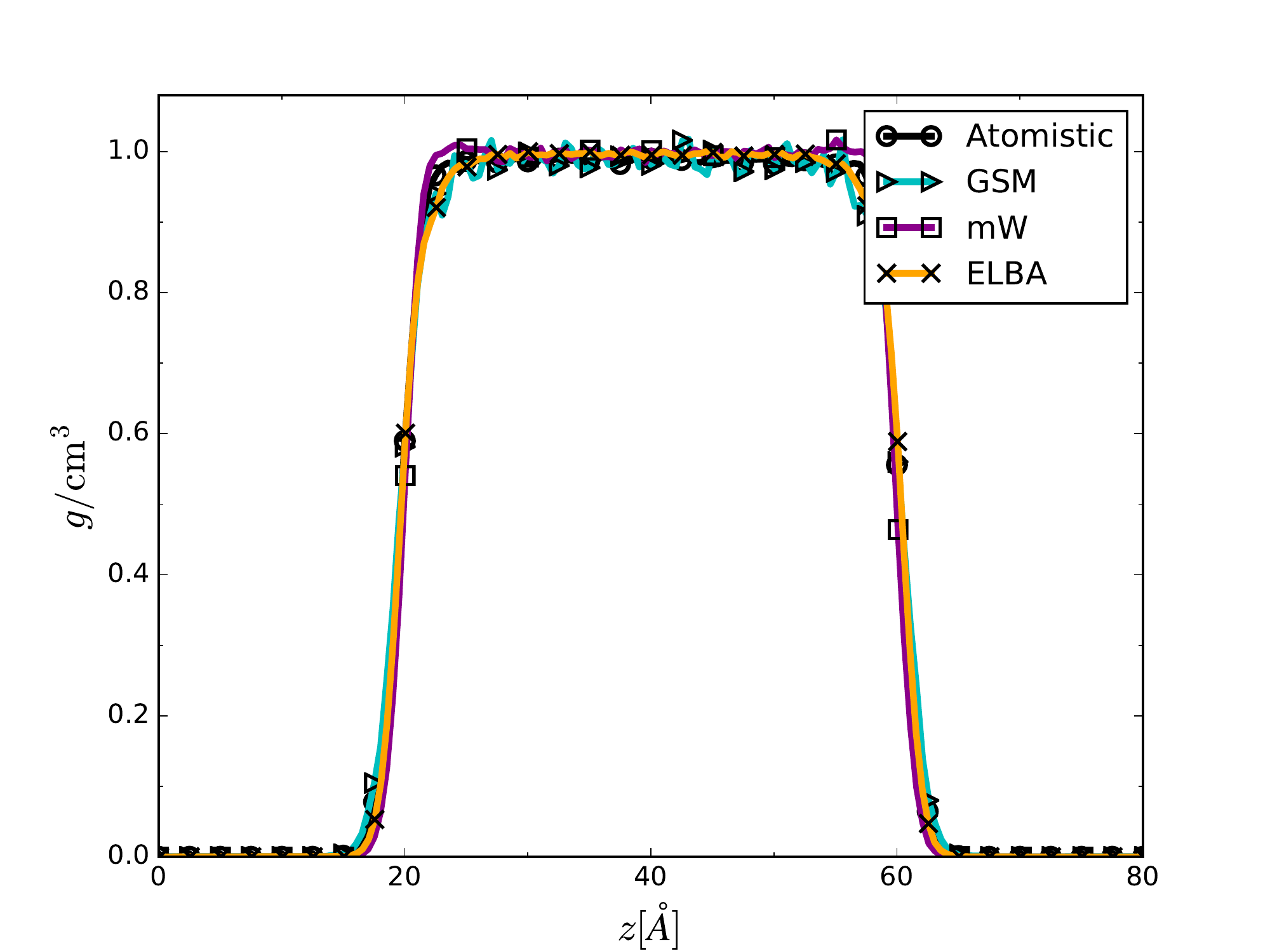}\\
    \caption{Density z-profile of the slab system at {\SI{300}{\kelvin}}.
    The center of the slab is at $z=4.0$\,{\AA}.}
\label{fig:slab}
\end{figure}

As shown in Fig.~\ref{fig:slab}, the density profiles of GSM, mW and the ELBA model are in good agreement with the atomistic model. As an example, for GSM the density matches well the atomistic model (\SI{0.9878}{\gram/\centi\meter\cubed} vs. \SI{0.9883}{\gram/\centi\meter\cubed} for SPC/E). 

\subsection{Transferability}

\begin{figure}[ht]
     \includegraphics[width=\columnwidth]{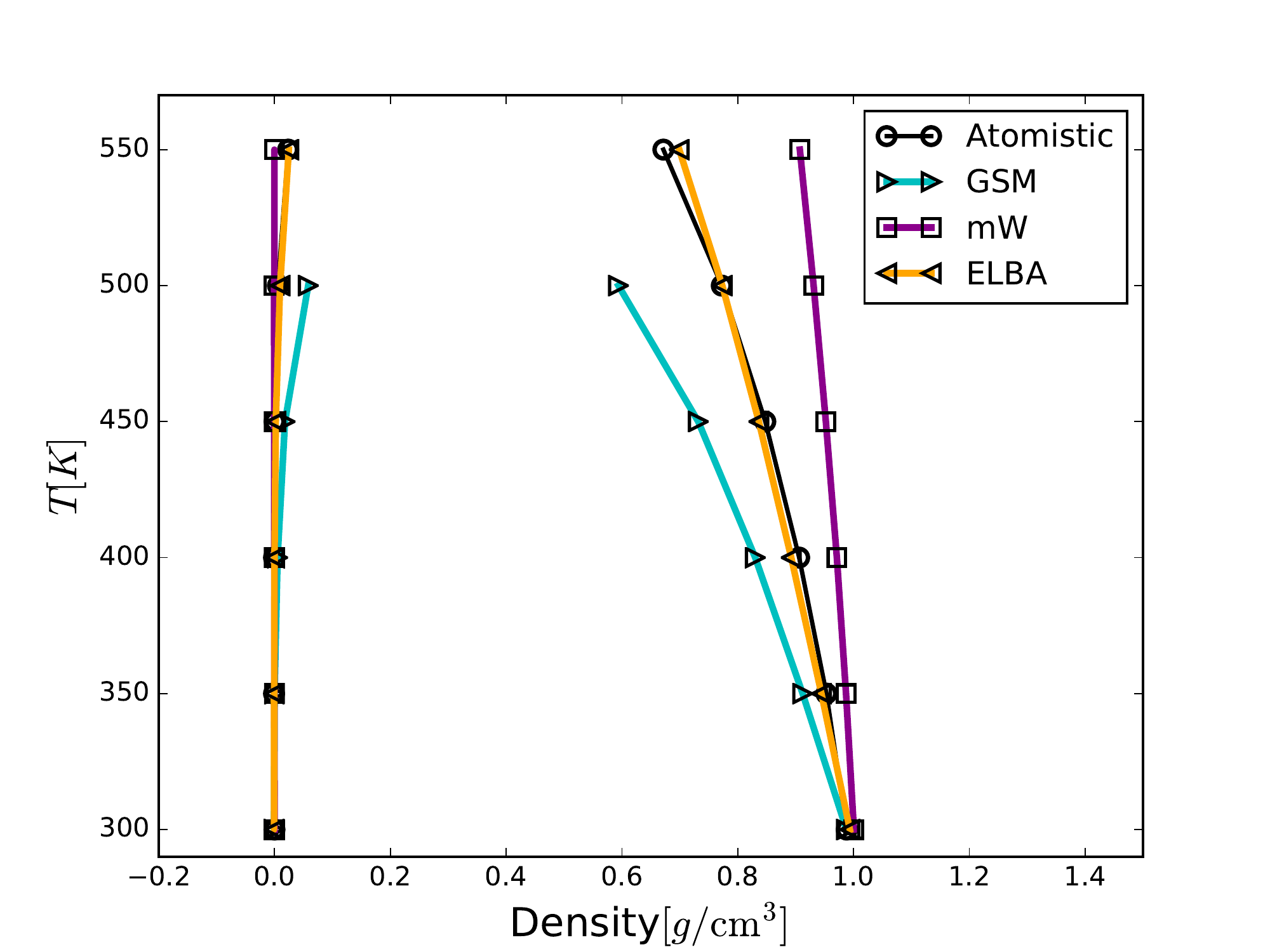}
     \caption{Vapor-liquid phase diagram for the GSM potential derived from slab simulations, SPC/E\cite{Berendsen1987}, mW\cite{Molinero2009} and ELBA\cite{Orsi2011,Orsi2013} models.
       Points near the critical point are not shown due to difficulty of finding densities in that region.}
\label{fig:phase_slab}
\end{figure}

Significant deviations in liquid density were observed when the GSM-slab (parametrized to density) and mW potentials were used in slab systems at higher temperatures ($T \geq$ \SI{350}{\kelvin}), Fig.~\ref{fig:phase_slab}. Water described by the GSM-slab potential evaporates much easier than the reference and has a much lower triple point. The exact position of the triple point proved to be impossible to determine with current methods. Meanwhile, the mW model remained liquid and stayed at a nearly constant density regardless of temperature. This can likely be explained by the short range nature of its potential, where a molecule in gas phase is too far away from any other molecules to contribute to the total potential energy of the system.

Furthermore, when ANC-1G and ANC-2G potentials parametrized to bulk $g(r)$ were simulated in slab systems, both evaporated and spread evenly through the whole volume at all attempted temperatures. This is due to the repulisive nature of the potentials as discussed in the previous section. While they are suitable for use at the temperature and mean density they were parametrized at, these potentials, as well as the GSM and mW potentials, appear not to be transferable to other thermodynamic conditions. These results suggest that a transferable water model needs to be parametrized not only against multiple properties (density or $g(r)$, diffusion coefficient, tetrahedral order parameter, etc.), but possibly also at different state points (temperature, pressure) simultaneously. One possibility is to interpolate between such state points. Addressing this in detail is, however, beyond the current study.

\section{Conclusions} \label{section:conclusions}

An ideal water model should be both fast and accurate. However, as we coarse-grain the system to make the simulations faster, we also lose degrees of freedom, complexity of the potential, and (at least some) descriptive power of the model. This is exemplified by the ANC-1G and GSM parametrizations in the bulk system, where the functional forms of the potentials prevent accurate reproduction of $g(r)$ even though $g(r)$ was the target property for parameter optimization. This necessitates a more complex two-body potential, often one without an explicit functional form, like the one obtained from IBI.

While such potentials are unique for each state point\cite{Henderson1974,Johnson2007}, they change with thermodynamic variables like temperature and pressure. This is the transferability problem\cite{Johnson2007}: potentials parametrized for one state point are often poor representations of other state points. This is exemplified in the slab systems by the density profile of the of GSM and the complete evaporation of the ANC-1G and ANC-2G potentials. All existing {CG} models suffer from this to some degree and often involve a compromise between quantitatively reproducing the general trend over many states and reproducing any given state exactly.

Furthermore, even with a two-body effective potential exactly reproducing the $g(r)$ at a state point, it is impossible to recapture the thermodynamic properties of an underlying multi-body potential\cite{Johnson2007}, an effect known as representability. In water, the multi-body contribution arises from interactions involving non-uniform charge distribution, which are ignored in single-particle CG models. Representability problems can also be observed in anisotropic systems being modeled with isotropic potentials\cite{Louis2011}. Even though it is anisotropic, the point dipole introduced in the GSM is still a two-body potential. It is not enough to obtain accurate thermodynamic properties of water, so the potential needs to include either higher orders of the multipole expansion or a true multi-body contribution, likely one that explicitly incorporates hydrogen bonding. Therefore, in subsequent works, we will expand the GSM potential to include higher order moments and explore a parametrization method aimed to simultaneously fit 
physical properties at multiple state points. 
{
The best-fit parameters found here are given below in Appendix A.
}

\begin{acknowledgments}
T.R.-L. acknowledges the Consejo Nacional de Ciencia y Tecnolog\'ia (CONACyT), Mexico for financial support, M.K. was supported by Natural Sciences and Engineering Research Council of Canada (NSERC). Computational resources were provided by SharcNet [www.sharcnet.ca], WestGrid [www.westgrid.ca], and Compute Canada.
\end{acknowledgments}

\appendix

{
\section{Best-fit parameters}
The best-fit parameters for the ANC-1, ANC-2G, and GSM models are presented in Tables~\ref{tbl:bulk} and \ref{tbl:slab}. 
}

\begin{table}[h]
\caption{Values of the best-fit parameters obtained for ANC-1G and ANC-2G models 
(Eqs.~\ref{eq:upair_anc} and \ref{eq:gaussian}) 
with $g(r)$ as the target property}
\label{tbl:bulk}
\centering
\begin{tabular}{llll}
\hline 
 & ANC-1G & ANC-2G & GSM \\ 
\hline 
$\epsilon[\si{kcal/mol}]$ & 0.191 & 0.190 & 0.328 \\ 
$\delta_{\rm{m}}[\si{\angstrom}]$ & 2.89 & 2.90 & 3.24 \\ 
$s$ & 0.594 & 0.595 & 0.886 \\ 
$h_1[\si{kcal/mol}]$ & 1.78 & 1.80 & --\\ 
$p_1[\si{\angstrom}]$ & 2.93 & 2.92 & -- \\ 
$q_1[\si{\angstrom}]$ & 0.686 & 0.692 & -- \\ 
$h_2[\si{kcal/mol}]$ & -- & 0.102 & -- \\ 
$p_2[\si{\angstrom}]$ & -- & 5.53 & -- \\ 
$q_2[\si{\angstrom}]$ & -- & 0.368 & -- \\ 
$\mu[\si{D}]$ & -- & -- & 3.58 \\
\hline 
\end{tabular} 
\end{table}

\begin{table}[h]
\caption{Values of the best-fit parameters obtained for GSM model
(Eq.~\ref{eq:upair_gsm} ) 
with density $z$-profile as the target property}
\label{tbl:slab}
\centering
\begin{tabular}{lll}
\hline
 															& GSM \\ 
\hline
$\epsilon[\si{kcal/mol}]$  					& 0.549 \\  
$\delta_{\rm{m}}[\si{\angstrom}]$	& 3.32 \\ 
$s$														& 0.928 \\ 
$\mu[\si{D}]$										& 2.51 \\ 
\hline
\end{tabular} 
\end{table}


%

\end{document}